\documentclass[twocolumn,amsmath,amssymb,aps, prl,
showkeys,showpacs, superscriptaddress, floatfix] {revtex4-2}
\usepackage{graphicx}
\usepackage[usenames,dvipsnames]{color}
\usepackage{bm}
\usepackage[colorlinks=true,breaklinks=true,allcolors=blue]{hyperref}
\usepackage{dcolumn}
\usepackage{slashed}
\usepackage{mathtools}
\usepackage{multirow}
\usepackage{comment}
\usepackage{amsmath}

\usepackage{tikz}
\usetikzlibrary{shapes.geometric}   
\usetikzlibrary{calc,decorations.pathmorphing}
\usetikzlibrary{decorations.pathreplacing}
\usetikzlibrary{decorations.markings,arrows.meta}
\usetikzlibrary{shapes}
\usepackage{pgf}
\usetikzlibrary{calc}

\newcommand{\vvv}{
  \coordinate (base) at (0, .8);
  \node[fill, circle, minimum size = 2mm, inner sep = 0] at (0, 0) (a) {};
  \node[fill, circle, minimum size = 2mm, inner sep = 0] at (3, 0) (c) {};
  \node[fill, circle, minimum size = 2mm, inner sep = 0] at (1.5, 2.6) (b) {};
}
\allowdisplaybreaks

\newcommand{\QD}{\mathrm{QD}}

\newcommand{\vertices}[1]{
\draw (0, 0) node[fill, circle, minimum size = 0.5mm, inner sep = 0]{};
  \draw (3, 0) node[fill, circle, minimum size = 0.5mm, inner sep = 0]{};
  \draw (0, 3) node[fill, circle, minimum size = 0.5mm, inner sep = 0]{};
  \draw (3, 3) node[fill, circle, minimum size = 0.5mm, inner sep = 0]{};
}

\usetikzlibrary{shapes.misc}
\tikzset{cross/.style={cross out, draw=black, thick, fill=none, minimum size=2*(#1-\pgflinewidth), inner sep=0pt, outer sep=0pt}, cross/.default={3pt}}

\allowdisplaybreaks

\begin{document}
\title{Exact solution of three-point functions in critical loop models}
\author{Morris Ang}
\email{moang@ucsd.edu}
\affiliation{Department of Mathematics, UC San Diego, USA}
\author{Gefei Cai}
\email{caigefei1917@pku.edu.cn}
\affiliation{Beijing International Center for Mathematical Research, Peking University, PR China}
\author{Jesper Lykke Jacobsen}
\email{jesper.jacobsen@phys.ens.fr}
\affiliation{Laboratoire de Physique de l'École Normale Supérieure, ENS, Université PSL, CNRS, Sorbonne Université, Université Paris Cité, Paris, France}
\author{\\ Rongvoram Nivesvivat}
\email{rongvoramnivesvivat@gmail.com}
\affiliation{New York University Abu Dhabi, Abu Dhabi, United Arab Emirates}
\author{Paul Roux}
\email{paul.roux@phys.ens.fr}
\affiliation{Laboratoire de Physique de l'École Normale Supérieure, ENS, Université PSL, CNRS, Sorbonne Université, Université Paris Cité, Paris, France}
\affiliation{Institut de physique théorique, CEA, CNRS,
Université Paris-Saclay}
\author{Xin Sun}
\email{xinsun@bicmr.pku.edu.cn}
\affiliation{Beijing International Center for Mathematical Research, Peking University, PR China}
\author{Baojun Wu}
\email{wubaojunmathe@outlook.com}
\affiliation{Beijing International Center for Mathematical Research, Peking University, PR China}
 
\begin{abstract}
We propose an exact formula  for three-point functions on the sphere in critical loop models with primary fields  $V_{(r,s)}$  characterized by $2r$ legs and a parameter \(s\) that describes diagonal fields for $r=0$ and the momentum of legs for $r>0$. We demonstrate its validity in three ways: a transfer-matrix study of the lattice model, the conformal bootstrap method for 4-point functions,  and a probabilistic method based on conformal loop ensemble and Liouville quantum gravity. This work provides a crucial missing piece for solving critical loop models and reveals a deep unity between three fundamental approaches to  2D statistical physics: transfer matrix, conformal field theory,  and probability theory.
\end{abstract}
\maketitle 

\noindent{\bf Introduction.}
Two-dimensional (2D) critical loop models occupy a unique position in theoretical physics. They describe the universal properties of many statistical systems, from the \(Q\)-state Potts model to the O(\(n\)) model~\cite{nie82, fsz87}, while simultaneously providing a geometric framework for conformal field theory (CFT)~\cite{belavin1984infinite, francesco2012conformal} and serving as a central object of study in probability theory through Schramm-Loewner Evolution (SLE) and Conformal Loop Ensembles (CLE)~\cite{msw-non-simple, msw-cle-lqg}. Understanding these models in full has been a long-standing goal, with significant progress made on critical exponents~\cite{nie82,cardy1992critical,NQSZ-backbone} and partition functions~\cite{fsz87,Car06,DJS2009}.

A complete solution requires knowledge of the correlation functions. On the sphere, one-point functions vanish, and two-point functions are determined by critical exponents, up to field normalizations.
The next outstanding object
for critical loop models is then the sphere three-point function of primary fields of the form 
$V_{(r,s)}$. The parameter $r \ge 0$ is half the number of open loop segments (legs)  inserted by $V_{(r,s)}$, while $s$ gives their 
momentum or, for $r=0$, modifies the fugacity of loops that wind around the insertion point \cite{gjnrs23, gnjrs21}. 

In this Letter, we report a general exact formula~\eqref{C123}  for the three-point functions of these primary fields in critical loop models, which  was proposed in~\cite{JNRR25} and inspired by~\cite{acsw21,nrj23}. 
Previously known results~\cite{DF84,delfino2010three, ikhlef2016three} concern diagonal operators, and have the form of the imaginary DOZZ formula~\cite{Schomerus2003,zamolodchikov2005three, Kostov-DOZZ,krv-dozz} that is crucial to Liouville theory and related CFTs.
Expressed using the Barnes double gamma function, our formula~\eqref{C123} goes significantly beyond the imaginary DOZZ formula and captures much more geometric observables. In particular, we obtain the three-point function for the two-leg operator, which encompasses the probability that three points are on the same loop. In the rest of the Letter, we present our result and describe the three complementary methods coming from transfer matrix, CFT,  and probability theory.

\smallskip

\noindent{\bf Critical loop models.}
The CFT of critical loop models is parametrized by a coupling  constant $\beta^2$
such that ${\rm Re}\, \beta^2 > 0$.
We take generic values of $n$ and $Q=n^2$; our main interest is the interval $\beta^2 \in [\frac12, \frac32]$ corresponding to $0 \le n \le 2$.
The torus partition function can be computed \cite{fsz87} from the Coulomb gas (CG) method \cite{nie82}. It determines the central charge $c$ and the loop fugacity $n$,
\begin{equation}
\label{cn}
c = 13 - 6\beta^2 - 6\beta^{-2} \,, \qquad n = -2\cos(\pi\beta^2) \,,
\end{equation}
of the underlying O(\(n\)) or $Q$-state Potts model ($Q=n^2$).
It also determines the {\em spectrum}, a set of primary fields \(V_{(r,s)}\) with left and right conformal dimensions \((\Delta, \bar\Delta) = (\Delta_{(r,s)}, \Delta_{(r,-s)})\) which can be written as
\begin{equation}
\label{Drs}
 \Delta_{(r,s)} = P_{(r,s)}^2 - P_{(1,1)}^2 \,, \quad P_{(r,s)} = \frac12(r \beta-s \beta^{-1}) \,.
\end{equation}
The parameters take values \(r \in \frac{1}{2}\mathbb{N}\) and \(s \in \mathbb{C}\), subject to the constraint that the conformal spin \(rs \in \mathbb{Z}\) \cite{nrj23}. If $r=0$ and $s+\beta^2 \in \mathbb{N}^*$ the field is degenerate at level $s$. In geometrical terms, a field \(V_{(r,s)}\) with \(r>0\) inserts \(2r\) open loop segments, each taking a phase ${\rm e}^{i \pi s}$ when winding around the insertion point, while a diagonal field \(V_{(0,s)}\) modifies the weight of loops that surround it, replacing $n$ by \(w =2\cos(\pi s)\)~\cite{gjnrs23}.


\smallskip

\noindent{\bf Three-point functions.}
Global conformal invariance determines the coordinate dependence of two- or three-point correlation functions~\cite{belavin1984infinite}. 
In particular, using the short-hands $i = (r_i,s_i)$ and $z_{ij} = z_i - z_j$, the three-point function $\left\langle V_1(z_1) V_2(z_2) V_3(z_3) \right\rangle$ on the sphere reads
\begin{equation*}
 \frac{C_{123}}{|z_{12}|^{2(\Delta_1+\Delta_2-\Delta_3)} |z_{13}|^{2(\Delta_1+\Delta_3-\Delta_2)} |z_{23}|^{2(\Delta_2+\Delta_3-\Delta_1)}} \,,
\end{equation*}
where it remains to determine the structure constant $C_{123}$. Our main result is the following exact formula
\begin{equation}
\label{C123}
C_{123} = \hspace{-0.4cm}\prod_{\epsilon_1,\epsilon_2,\epsilon_3=\pm} \hspace{-0.4cm}\Gamma_{\beta}^{-1}\big(\tfrac{\beta+\beta^{-1}}{2}+\tfrac{\beta}{2}\big| \sum_i \epsilon_i r_i\big|+\tfrac{\beta^{-1}}{2}\sum_i \epsilon_i s_i\big) \,.
\end{equation}
The Barnes double gamma function $\Gamma_\beta$ is defined by its shift relations~\cite{sm}.
Eq.~\eqref{C123} was originally  used as an ansatz to solve the crossing symmetry for four-point functions~\cite{hjs20, nrj23}, but we propose it here as the exact expression for all three-point functions.
It is defined for all parameters satisfying the constraints
\begin{equation}
 r_1+r_2+r_3 \in \mathbb{N} \,, \qquad r_i s_i \in \mathbb{Z} \,,
\label{sumr}
\end{equation}
which guarantee that $\left\langle V_1(z_1) V_2(z_2) V_3(z_3) \right\rangle$ is non-zero and single-valued \cite{nrj23}.
The three-point functions of degenerate fields and diagonal fields $V_{(0,s)}$ were previously known, from the CG \cite{DF84} and imaginary Liouville theory \cite{delfino2010three, ikhlef2016three} respectively; both are special cases of \eqref{C123}. Indeed, setting $r_1=r_2=r_3=0$ in~\eqref{C123} recovers the imaginary DOZZ formula~\cite{zamolodchikov2005three, krv-dozz, Kostov-DOZZ}.

\smallskip

\noindent{\bf Combinatorial map.}
A key insight from the lattice construction is that this structure constant has a direct combinatorial interpretation~\cite{gjnrs23, gnjrs21}. It can be associated to a unique \textit{combinatorial map} connecting the three punctures on the sphere, where each puncture labeled by \((r_i, s_i)\) contributes \(2r_i\) legs. The map is formed by connecting these legs in pairs, respecting the cyclic order around each puncture. Here are examples with $(3,3,2)$, $(4,1,1)$ and $(6,2,0)$ legs:
\begin{align}
  \label{fig:cmaps}
  \begin{tikzpicture}[baseline=(base), scale = .5]
    \vvv;
    \draw (a) -- (c);
    \draw (b) -- (c);
    \draw (a) to [out = 75, in = -135] (b);
    \draw (a) to [out = 45, in = -105] (b);
  \end{tikzpicture}
  \qquad 
  \begin{tikzpicture}[baseline=(base), scale = .5]
    \vvv;
    \draw (a) -- (c);
    \draw (a) -- (b);
    \draw (a) to [out = -15, in = -90] (3.5, 0) to [out = 90, in = 15] (a);
  \end{tikzpicture}
  \qquad 
  \begin{tikzpicture}[baseline=(base), scale = .5]
    \vvv;
    \draw (a) to [out = 75, in = -135] (b);
    \draw (a) to [out = 45, in = -105] (b);
    \draw (a) to [out = -15, in = -90] (3.5, 0) to [out = 90, in = 15] (a);
    \draw (a) to [out = -30, in = -90] (3.8, 0) to [out = 90, in = 30] (a);
  \end{tikzpicture}
\end{align}
The last two examples show that self-contractions are allowed only if there is an operator inside the enclosure. The first of \eqref{sumr} is just the hand-shake lemma of the map.
On the sphere with punctures at $z_i$ we can thus understand \(\langle V_1(z_1) V_2(z_2) V_3(z_3)\rangle\) physically as the partition function of the loop model conditioned by this specific connectivity pattern, with weights $n$ for each contractible loop and $w_i$ for each non-contractible loop around $z_i$ (possible only if $r_i = 0$). The phases ${\rm e}^{i \pi s_i}$ will be discussed below.


This geometric picture provides a natural bridge between the lattice model and the CFT. Eq.~\eqref{C123} is the analytical expression for the structure constant corresponding to this combinatorial map. To obtain a quantity that is invariant under field renormalizations \(V_{(r,s)}\to \lambda_{(r,s)} V_{(r,s)}\), we work with the normalized ratio
\begin{equation}
\omega_{123} = C_{123} \sqrt{\frac{C_{000}}{C_{011}C_{022}C_{033}}} \,,
\label{om123}
\end{equation}
where the subscript \(0\) denotes the (degenerate) identity field \(V_{(0,1-\beta^2)}\)~\cite{nrj23}.
The factors in the denominator are two-point structure constants, and $C_{000}$ is the partition function.  The following sections provide compelling evidence for this proposal from three independent frameworks: numerical transfer matrix calculations, conformal bootstrap, and conformal loop ensembles.

\begin{figure}[htbp]
  \centering
  \includegraphics[width=0.48\textwidth]{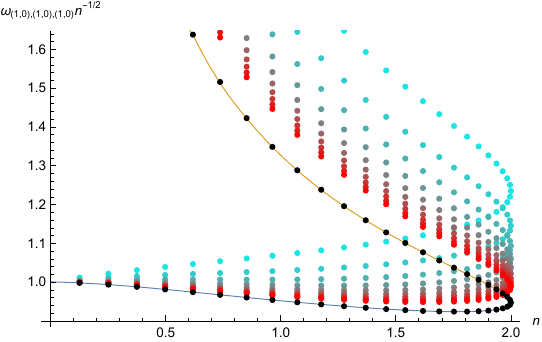}
  \caption{The colored points are numerical results for $\omega_{(1,0)(1,0)(1,0)}(L) n^{-1/2}$ in the $O(n)$ model, as functions of $n$, for finite sizes ranging from $L=4$ (cyan) to $L=13$  (red). The black points are $L \to \infty$ extrapolations from $8 \le L \le 13$. The continuous line is the analytical result \eqref{om123} with \eqref{C123}, with a lower branch (blue) for the dense phase ($\tfrac12 \le \beta^2 \le 1$), and an upper branch (orange) for the dilute phase ($1 \le \beta^2 \le \tfrac32$). 
  }
  \label{fig:3ptOn222}
\end{figure}

\noindent{\bf Transfer matrix.}
We choose as a lattice discretization the O($n$) loop model on the square lattice \cite{BN1989} (the cognate O($n$) model on the hexagonal lattice \cite{nie82} is in the same universality class). Its local weights and vertex configurations are:

\begin{align}
  \begin{tikzpicture}[baseline={([yshift=-3pt]current bounding box.center)}, scale = .3]
    \draw[gray] (0,-1) -- (0,1);
    \draw[gray] (-1,0) -- (1,0);
    \node[below] at (0,-1) {$\rho_1$};
  \end{tikzpicture} \quad
  \begin{tikzpicture}[baseline={([yshift=-3pt]current bounding box.center)}, scale = .3]
    \draw[gray] (0,-1) -- (0,1);
    \draw[gray] (-1,0) -- (1,0);
    \draw[very thick,blue] (-1,0) to (-0.3,0) arc (-90:0:0.3) to (0,1);
    \node[below] at (0,-1) {$\rho_2$};
  \end{tikzpicture} \quad
  \begin{tikzpicture}[baseline={([yshift=-3pt]current bounding box.center)}, scale = .3]
    \draw[gray] (0,-1) -- (0,1);
    \draw[gray] (-1,0) -- (1,0);
    \draw[very thick,blue] (0,-1) to (0,-0.3) arc (180:90:0.3) to (1,0);
    \node[below] at (0,-1) {$\rho_3$};
  \end{tikzpicture} \quad
  \begin{tikzpicture}[baseline={([yshift=-3pt]current bounding box.center)}, scale = .3]
    \draw[gray] (0,-1) -- (0,1);
    \draw[gray] (-1,0) -- (1,0);
    \draw[very thick,blue] (0,-1) to (0,-0.3) arc (0:90:0.3) to (-1,0);
    \node[below] at (0,-1) {$\rho_4$};
  \end{tikzpicture} \quad
  \begin{tikzpicture}[baseline={([yshift=-3pt]current bounding box.center)}, scale = .3]
    \draw[gray] (0,-1) -- (0,1);
    \draw[gray] (-1,0) -- (1,0);
    \draw[very thick,blue] (0,1) to (0,0.3) arc (180:270:0.3) to (1,0);
    \node[below] at (0,-1) {$\rho_5$};
  \end{tikzpicture} \quad
  \begin{tikzpicture}[baseline={([yshift=-3pt]current bounding box.center)}, scale = .3]
    \draw[gray] (0,-1) -- (0,1);
    \draw[gray] (-1,0) -- (1,0);
    \draw[very thick,blue] (-1,0) to (1,0);
    \node[below] at (0,-1) {$\rho_6$};
  \end{tikzpicture} \quad
  \begin{tikzpicture}[baseline={([yshift=-3pt]current bounding box.center)}, scale = .3]
    \draw[gray] (0,-1) -- (0,1);
    \draw[gray] (-1,0) -- (1,0);
    \draw[very thick,blue] (0,-1) to (0,1);
    \node[below] at (0,-1) {$\rho_7$};
  \end{tikzpicture} \quad
  \begin{tikzpicture}[baseline={([yshift=-3pt]current bounding box.center)}, scale = .3]
    \draw[gray] (0,-1) -- (0,1);
    \draw[gray] (-1,0) -- (1,0);
    \draw[very thick,blue] (-1,0) to (-0.3,0) arc (-90:0:0.3) to (0,1);
    \draw[very thick,blue] (0,-1) to (0,-0.3) arc (180:90:0.3) to (1,0);
    \node[below] at (0,-1) {$\rho_8$};
  \end{tikzpicture} \quad
  \begin{tikzpicture}[baseline={([yshift=-3pt]current bounding box.center)}, scale = .3]
    \draw[gray] (0,-1) -- (0,1);
    \draw[gray] (-1,0) -- (1,0);
    \draw[very thick,blue] (0,-1) to (0,-0.3) arc (0:90:0.3) to (-1,0);
    \draw[very thick,blue] (0,1) to (0,0.3) arc (180:270:0.3) to (1,0);
    \node[below] at (0,-1) {$\rho_9$};
  \end{tikzpicture}
  \nonumber
\end{align}
For a certain trigonometric choice of the loop weight $n$ and the vertex weights $\rho_i$ \cite{BN1989} this model is known \cite{nie82,fsz87} to converge to the CFT described by \eqref{cn}--\eqref{Drs} in the limit of a large lattice.
We now describe the lattice computation of $C_{123}$ in the case of all $r_i \neq 0$ and no enclosures, such as the first of \eqref{fig:cmaps}.
First, discretize the three-punctured sphere by an $L \times 2M$ cylinder:
  \begin{align}
    \hspace{-.5cm}
    \label{fig:cyl-lat}
    \begin{tikzpicture}[scale=0.5, rotate=-90,baseline=(current bounding box.center),
      oriented/.style={postaction={decorate},decoration={markings,
          mark=at position #1 with {\arrow{>}}}}
      ]
      \foreach \i in {1, 2, 3, 4, 5, 6, 7, 8, 9, 10}
      \draw (2, \i-0.5) arc[start angle=0, end angle=180, x radius=2cm, y radius=-0.2cm]; 
      \pgfmathsetmacro{\L}{6}
      \foreach \i in {0, 1, 2, 3, 4, 5, 6} {
        \pgfmathsetmacro{\t}{(6-\i) * 180 / \L}
        \pgfmathsetmacro{\x}{2 * cos(\t)}
        \pgfmathsetmacro{\y}{-0.2 * sin(\t)}
        \draw (\x, \y) -- ++ (0, 10);
      }
      \draw[red, postaction={decorate}, decoration={markings, mark=at position 0.75 with {\arrow[scale=1.3]{Stealth[reversed]}}}]
      (0, 10) ellipse[x radius=2cm, y radius=0.2cm];
      \node[right, red] at (2.5, 9.5) {$R_3$};
      \draw[red, red, postaction={decorate}, decoration={markings, mark=at position 0.5 with {\arrow[scale=1.3]{Stealth[reversed]}}}] (2, 0) arc[start angle=0, end angle=180, x radius=2cm, y radius=-0.2cm];
      \node[right, red] at (2.5, -0.5) {$R_1$}; 
      \node[left] at (-2, 0) {$_{t=-M}$}; 
      \draw[gray, dashed] (2, 5) arc[start angle=0, end angle=180, x radius=2cm, y radius=-0.2cm]; 
      \node[left] at (-2.5, 5.5) {$_{t=0}$}; 
      \node[left] at (-2, 12) {$_{t=M}$}; 
      \draw[red, dashed] (2, 0) arc[start angle=0, end angle=180, x radius=2cm, y radius=0.2cm]; 
      \draw[red, red, postaction={decorate}, decoration={markings, mark=at position 0.75 with {\arrow[scale=1.3]{Stealth[reversed]}}}] (0, 4.8) ellipse[x radius=1.2cm, y radius=5pt];
      \node[red, right] at (2.5, 4.3) {$R_2$};
      \node[left=3pt] (0, -0.2) {$L$};
    \end{tikzpicture}
  \end{align}
The punctures are the regions $R_i$: the two cylinder ends and a middle region $R_2$ of $r_2$ broken edges. The transfer matrix grows the lattice by evolving the time $t$ from left to right. It acts on states describing the partially constructed system of blue curves, as seen to the left of a given $t$: a set of non-intersecting pairwise connection arcs and defect lines going back to $R_1$ (and also $R_2$ if $t>0$). These are {\em link patterns} that span {\em standard modules}, technically representations of the dilute unoriented Jones-Temperley-Lieb algebra \cite{JNRR25}. Under the time evolution the arcs close so as to form loops, or are used for backtracking by the defects. The defects are labelled by their point of origin ($2r_1$ points are chosen in $R_1$ and the other $L-2r_1$ are left empty), and we impose constraints so that the connections made between the three $R_i$ realize the combinatorial map. Since the combinatorial map is defined up to cyclic rotations of the punctures, it can be realized by the labelled defects in $N = \prod(2r_i)$ ways. Pick one of these as a reference and relate the others to it by a triple of rotations $(\sigma_1,\sigma_2,\sigma_3)$, where $\sigma_i \in \mathbb{Z}_{2r_i}$. We then compute the $N$ conditioned partition functions $Z_{\sigma_1,\sigma_2,\sigma_3}$ and combine them into
\begin{align}
 Z_{123}(L,M) = 
  \sum_{\sigma_i\in\mathbb{Z}_{2 r_i}} {\rm e}^{i \pi (s_1 \sigma_1 + s_2 \sigma_2 + s_3 \sigma_3)} Z_{\sigma_1, \sigma_2, \sigma_3}\ \,.
\end{align}
Taking $M \gg L$ we obtain (to within numerical precision)
\begin{equation}
 C_{123}(L) = \lim_{M \to \infty} \frac{Z_{123}(L,M)}{Z_{000}(L,M)} \,,
\end{equation}
a finite-$L$ approximation to $C_{123}$.
Finally we form $\omega_{123}(L)$ from \eqref{om123} and extrapolate to the continuum limit, $L \to \infty$. A representative result is shown in Fig.~\ref{fig:3ptOn222}. The agreement with \eqref{C123} is excellent.

\smallskip

\noindent{\bf Conformal bootstrap.}
In CFT, the assumed existence and associativity of the operator-product
expansion (OPE) of local operators can be translated into the requirement of crossing symmetry of correlation functions. This also constrains the three-point functions, as we shall see. Consider first the crossing-symmetry of four-point functions,
\begin{align}
\sum_{V\in\mathcal{S}^{(s)}}
\hspace{-0.15cm}
d_{V}^{(s)}
 \begin{tikzpicture}[baseline=(B.base),  thick, scale = .1]
\draw (-1,2) node [above] {$_2$} -- (0,0) -- node [above] {$_V$} (4,0) -- (5,2) node [above] {$_3$};
\node (B) at (0, -.5){};
\draw (-1,-2) node [below] {$_1$} -- (0,0);
\draw (4,0) -- (5,-2) node [below] {$_4$};
\end{tikzpicture} 
\hspace{-0.09cm}
=
\hspace{-0.09cm}
\sum_{V\in\mathcal{S}^{(t)}}
\hspace{-0.15cm}
d_{V}^{(t)}
\begin{tikzpicture}[baseline=(B.base),  thick, scale = .1]
\draw (-2,1) node [above] {$_2$} -- (0,0) -- node [right] (B) {$_V$} (0,-4) -- (2,-5) node [below] {$_4$};
\draw (-2,-5) node [below] {$_1$} -- (0,-4);
\draw (0,0) -- (2,1) node [above] {$_3$};
\end{tikzpicture} 
\hspace{-0.09cm}
=
\hspace{-0.09cm}
\sum_{V\in\mathcal {S}^{(u)}}
\hspace{-0.15cm}
d_{V}^{(u)}
\begin{tikzpicture}[baseline=(B.base),  thick, scale = .1]
\draw (-2,1) node [above] {$_2$} -- (0,0) -- node [left](B) {$_V$} (0,-4) ;
\draw (0,0)-- (2,-5) node [below] {$_4$};
\draw (-2,-5) node [below] {$_1$} -- (0,-4);
\draw (0,-4) -- (2,1) node [above] {$_3$};
\end{tikzpicture} 
\label{3ch}
 ,
\end{align}
expressing the 3 different orders of performing the OPE, called the $x\in\{s, t, u\}$ channels. The unknowns in \eqref{3ch} are the structure constants $d_{V}^{(x)}$, while the diagrams represent conformal blocks, which are completely known (including logarithmic parts) for critical loop models \cite{nr20}. 
For $s_0 \in \mathbb{C} \setminus (2\mathbb{N}-\beta^2)$, write the spectrum as a disjoint union of degenerate, diagonal and non-diagonal fields:
\begin{eqnarray}
 \mathcal{S}^{\text{loop}}_{s_0} &=& \left\{ V^{\phantom{d}}_{(0, s)}\right\}_{s+\beta^2\in2\mathbb{N}+1}
 \bigcup 
\left\{ V^{\phantom{d}}_{(0,s)}\right\}_{s \in s_0 + 2 \mathbb{Z}} \nonumber \\
 & & \quad 
\bigcup \left\{V^{\phantom{d}}_{(r, s)}\right\}_{r\in \frac{\mathbb{N}^*}{2},s \in  \frac{\mathbb{Z}}{r}}
\ .
\label{spec}
\end{eqnarray}
The $d_{V}^{(x)}$ obey a degenerate-shift equation \cite{mr17}, due to the degenerate fields, so blocks in \eqref{3ch} differing by $2\mathbb{Z}$ on the second Kac label can be resummed into a single {\em interchiral conformal block} \cite{hjs20}. Hence $s$ in \eqref{spec} can be considered modulo $2 \mathbb{Z}$, and the reexpression of \eqref{3ch} in terms of interchiral blocks has fewer unknowns.

We solve all of \eqref{3ch} simultaneously, supposing that the spectrum $\mathcal{S}^{(x)}$ of exchanged fields $V$ is $\mathcal{S}^{\text{loop}}_{P_x}$, with only one interchiral family of degenerate or diagonal fields. This is an infinite linear system with a finite number of solutions \cite{gjnrs23}. The analytical expressions of $d_{V}^{(x)}$ are found \cite{nrj23} to be a product of two parts: Barnes' double Gamma functions and rational fractions in the loop fugacities. The former part is completely known, whereas the latter can be determined on a case-by-case basis \cite{nrj23}.
Physically,  the crossing-symmetric solutions to \eqref{3ch} describe the probabilities of loop configurations underlaid by combinatorial maps, obtained by connecting loop segments from each inserted local operator, as discussed in \eqref{fig:cmaps}. But unlike three-point functions, higher-point functions could have more than one associated combinatorial map, e.g.\
the 6 solutions of \eqref{3ch} for
$\left<V_{(1,0)}(z_1)V_{(1,0)}(z_2)V_{(1,0)}(z_3)V_{(1,0)}(z_4)\right>$ are in one-to-one
correspondence with the 6 combinatorial maps
\begin{align}
&
 \begin{tikzpicture}[baseline={([yshift=-0.5pt]current  bounding  box.center)}, scale = .2]
 \vertices{}
  \node[left] at (0, 0) {$z_2$};
 \node[left] at (0, 3) {$z_1$};
 \node[right] at (3, 3) {$z_4$};
 \node[right] at (3, 0) {$z_3$};
  \draw (0, 0) -- (0, 3) to [out = -60, in = 60] (0, 0);
  \draw (3, 0) -- (3, 3) to [out = -60, in = 60] (3, 0);
    \end{tikzpicture}
\ , \qquad
 \begin{tikzpicture}[baseline=(current  bounding  box.center), scale = .2]
 \vertices{}
  \draw (0, 0) -- (3, 0) to [out = 150, in = 30] (0, 0);
  \draw (0, 3) -- (3, 3) to [out = 150, in = 30] (0, 3);
 \end{tikzpicture}
\ , \qquad
 \begin{tikzpicture}[baseline={([yshift=3.0pt]current  bounding  box.center)}, scale = .2]
 \vertices{}
  \draw (3, 0) -- (0, 3) to [out = -15, in = 105] (3, 0);
  \draw (0, 0) to [out = -30, in = -135] (3.5, -.5) to [out = 45, in = -60] (3, 3);
  \draw (0, 0) to [out = -60, in = -135] (4, -1) to [out = 45, in = -30] (3, 3);
 \end{tikzpicture}
 \ ,
  \nonumber
  \\
  &
 \begin{tikzpicture}[baseline={([yshift=3pt]current  bounding  box.center)}, scale = .2]
  \vertices{}
    \node[left] at (0, 0) {$z_2$};
 \node[left] at (0, 3) {$z_1$};
 \node[right] at (3, 3) {$z_4$};
 \node[right] at (3, 0) {$z_3$};
  \draw (0, 0) -- (3, 0) -- (3, 3) -- (0, 3) -- (0, 0);
 \end{tikzpicture}
\ , \qquad
 \begin{tikzpicture}[baseline={([yshift=3pt]current  bounding  box.center)}, scale = .2]
  \vertices{}
  \draw (0, 0) -- (3, 0) -- (0, 3) -- (3, 3);
  \draw (0, 0) to [out = -30, in = -135] (3.5, -0.5) to [out = 45, in = -60] (3, 3);
 \end{tikzpicture}
 \hspace{-0.25cm}
\ \ ,\qquad
 \begin{tikzpicture}[baseline={([yshift=3pt]current  bounding  box.center)}, scale = .2]
  \vertices{}
  \draw (0, 0) -- (0, 3) -- (3, 0) -- (3, 3);
  \draw (0, 0) to [out = -30, in = -135] (3.5, -0.5) to [out = 45, in = -60] (3, 3);
 \end{tikzpicture}
 \ .
 \label{diag}
\end{align}

\smallskip

Consider now $\left<V_{(1,0)}V_{(1,0)}V_{(1,0)}\right>$ that describes the probability of  a loop passing through the 3 points. From the first combinatorial map in \eqref{diag}, assuming the normalization $\langle V_{(1,0)}(z)V_{(1,0)}(0)\rangle
=|z|^{-4\Delta_{(1,0)}}$, we have
\begin{align}
 \begin{tikzpicture}[baseline={([yshift=-0.5pt]current  bounding  box.center)}, scale = .2]
 \vertices{}
  \node[left] at (0, 0) {$0$};
 \node[left] at (0, 3) {$z$};
 \node[right] at (3, 3) {$1$};
 \node[right] at (3, 0) {$\infty$};
  \draw (0, 0) -- (0, 3) to [out = -60, in = 60] (0, 0);
  \draw (3, 0) -- (3, 3) to [out = -60, in = 60] (3, 0);
    \end{tikzpicture} \hspace{-0.3cm}
    \overset{z\rightarrow0}{=}
|z|^{-4\Delta_{(1,0)}} + A |z|^{-2\Delta_{(1,0)}}+
 \ldots
 \ ,
\label{ex1}
\end{align}
where $A = d_{V_{(1,0)}}^{(s)}/d_{V_{(0,1-\beta^2)}}^{(s)}$. 
We conjecture that $A = (C_{(1,0)(1,0)(1,0)}/n)^2$, where $n$ comes from reinterpreting the combinatorial map as a loop.
Using the results of \cite{nrj23}, we then find that $C_{(1,0)(1,0)(1,0)}$
indeed agrees with~\eqref{C123} in this case.
The general result \eqref{C123} is similarly motivated by solutions to \eqref{3ch} for other well-chosen four-point functions.


\smallskip

\noindent{\bf The probabilistic method.}
The third approach to critical loop models comes from probability theory, where the continuum limit is constructed directly as a Conformal Loop Ensemble (CLE)~\cite{CN06-full, SheffieldCLE}. This construction depends on a parameter \(\kappa\in (8/3,8) \)
related to parameters in \eqref{cn} by \(\kappa = 4\beta^{-2}\).  
CLE$_\kappa$ with parameter $\kappa$ is a conformally invariant random collection of loops, each of which behaves as an SLE$_\kappa$ curve~\cite{Sc00}.   

In this framework, correlation functions correspond to probabilities of geometric events involving loops. For example, the three-point function 
\(\langle V_{(1,0)}V_{(1,0)}V_{(1,0)}\rangle\) is proportional to the probability that a single CLE loop passes through three points. Although this event has probability zero, the three-point function can be understood as a scaling limit of the probability that a loop passes through three small disks centered at the three points. The normalized ratio taken as in~\eqref{om123} is independent of the scaling factor. By~\cite{Sm01,CN06-full,GPS2013},  the Bernoulli site percolation on the triangular lattice converges to CLE$_6$ in a rather strong sense. Hence, 
for the dense O$(1)$ model on the hexagonal lattice, the normalized ratio~\eqref{om123} for \(\langle V_{(1,0)}V_{(1,0)}V_{(1,0)}\rangle\) defined by CLE$_6$  agrees with the continuum  limit of the lattice counterpart. 

The key to the probabilistic approach is the coupling of SLE/CLE and random surfaces~\cite{DS-KPZ-invent} in Liouville quantum gravity (LQG) pioneered by Sheffield~\cite{She16a}. First, LQG surfaces coupled with CLE inherit a rich integrable structure from random planar maps decorated with the O$(n)$ loop model, 
as demonstrated in~\cite{DMS14,msw-non-simple,msw-cle-lqg}. Second, Liouville CFT, the field theory that describes LQG, is itself integrable.
In particular, the three-point function on the sphere is given by the DOZZ formula~\cite{krv-dozz,DO94,ZZ96}.  
The overarching strategy of combining insights from both sources of solvability was initiated in~\cite{AHS-SLE-integrability} and has since been successfully applied to several problems~\cite{nolin2024,ang2024boundary,ars-annuli}.

We demonstrate this approach using the derivation of \(\langle V_{(1,0)}V_{(1,0)}V_{(1,0)}\rangle\) in the dilute O($n$) model. 
Let $b\in (0,1)$, $Q=b+ b^{-1}$, and $c_{\mathrm L}=1+6Q^2$ be the coupling parameter, the background charge, and the central charge of the Liouville theory.
In the aforementioned coupling, the central charge for the Liouville CFT governing the LQG surfaces and the central charge $c_{\beta}$ for the loop model as in \eqref{cn} are related by $c_{\rm L}+c_\beta=26$. Here 26 is the dimension of the critical bosonic string~\cite{polyakov1981quantum}, and the loop model acts as a conformal matter field~\cite{david-conformal-gauge,dk-qg}. When $c_{\mathrm L}=26$, the LQG surface describes the continuum limit of 
uniformly sampled random planar maps, which models pure 2D gravity. In general, LQG surfaces with $c_{\mathrm L}=26-c_\beta$ describe the continuum limit of random planar maps decorated with the O$(n)$ loop model. 
Let $\QD_{0,3}$ be the LQG disk with three marked boundary points, 
conformally parameterized by the unit disk~$\mathbb D$. 
Then the law of the conformal factor $\phi$ is given by the Liouville CFT on $\mathbb D$ with 
an insertion at the three marked points. The insertions are of the form $e^{b\phi}$ thanks to 
the Knizhnik-Polyakov-Zamolodchikov (KPZ) relation~\cite{KPZrelation,DS-KPZ-invent}.

\begin{figure}[htbp]
  \centering
  \includegraphics[width=0.48\textwidth]{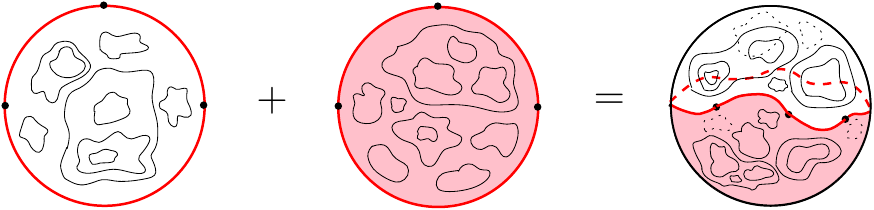}
  \caption{Illustration for the derivation of \(\langle V_{(1,0)}V_{(1,0)}V_{(1,0)}\rangle\).}\label{weld}
\end{figure}

Now we glue two independent copies of the CLE-decorated $\QD_{0,3}$ along their boundary. The resulting object is an LQG sphere decorated with CLE that has a loop passing through three marked points. See Figure~\ref{weld} for an illustration. 
The conformal factor $\phi$ describing the LQG sphere is given by Liouville theory on the sphere with three insertions of the form $e^{\alpha\phi}$. By the KPZ relation, $\alpha$ satisfies
\[
2d_{\mathrm L}(\alpha)+\Delta_{(1,0)}=2 \,,
\]
where
$d_{\mathrm L}(\alpha)= \frac{\alpha}{2}(Q-\frac{\alpha}{2})$ is the scaling dimension for the $e^{\alpha\phi}$-insertion, and 
$2-\Delta_{(1,0)}=1+\frac{\kappa}8$ is the fractal dimension of an SLE$_\kappa$ curve~\cite{Beffara_2008}. This yields $\alpha=b$.
Writing \(\langle e^{\alpha_1 \phi}e^{\alpha_2 \phi} e^{\alpha_3 \phi} \rangle\) as the three-point function of Liouville theory on the sphere, 
 the partition function of the right-hand side of Figure~\ref{weld} is then \(\langle V_{(1,0)}V_{(1,0)}V_{(1,0)}\rangle \times  \langle e^{b \phi}e^{b\phi} e^{b \phi} \rangle\). On the other hand, this quantity can be computed from $\QD_{0,3}$. 
Using the relation between the imaginary DOZZ formula at $c_\beta$ and the DOZZ formula at $c_{\rm L}$~\cite{Schomerus2003,zamolodchikov2005three,Kostov-DOZZ}, we arrive at an exact expression for $\langle V_{(1,0)}V_{(1,0)}V_{(1,0)}\rangle$, which agrees with~\eqref{C123} and~\eqref{om123}. See~\cite[Theorem 1.6]{acsw21}. This approach was also used in~\cite{acsw21} to derive the three-point function of diagonal operators $V_{(0,s)}$, which recovers the results from~\cite{delfino2010three,ikhlef2016three}. Furthermore, the approach was used in~\cite{CLWZ25} to derive an exact formula for the three-point function for the spin cluster in the $Q$-Potts model, which matches the numerical result from~\cite{DPSV13}.

\smallskip

\noindent{\bf Conclusions.}
Using three different methods, we gave evidence for the expression \eqref{C123} for the three-point structure constants. This subsumes previous results for degenerate \cite{DF84} and diagonal \cite{delfino2010three, ikhlef2016three} operators, and shows that correlation functions of legged operators $V_{(r,s)}$ with $r>0$ in critical loop models require a framework beyond the CG and imaginary Liouville approaches. Several directions remain open.
First, although we could extract three-point structure constants from
solutions to \eqref{3ch}, the four-point constants $d^{(x)}_V$ are not in general the product of two three-point constants. It was noticed in
\cite{nrj23} that there is in most cases an extra factor, a rational
function of the loop weights $n$ and $w_i$. Solving the four-point functions requires bringing this factor under analytical control. Second, in rational CFT the $C_{123}$ play the dual role of structure constants and operator-product expansion (OPE) coefficients. Here we lack a full understanding of fusion of $V_{(r,s)}$ operators. Third, tackling other geometries than the sphere requires further control of conformal boundary conditions and bulk-boundary fusion rules. We plan to come back to those issues in future work.

\smallskip
\noindent{\bf Acknowledgements.}
We thank Sylvain Ribault and Hubert Saleur for inspiring discussions. 
M.A. is  supported by DMS-2348201.
J.J. and P.R are supported by the French Agence Nationale de la Recherche (ANR) under grant ANR-21-CE40-0003 (project CONFICA). G.C., X.S. and B.W.\ are  supported by National Key R\&D Program of China (No.\ 2023YFA1010700) and the National Natural Science Foundation of China Grant (No. 12526204).

\bibliographystyle{apsrev4-1}
\bibliography{ref}

\begin{thebibliography}{48}%
\makeatletter
\providecommand \@ifxundefined [1]{%
 \@ifx{#1\undefined}
}%
\providecommand \@ifnum [1]{%
 \ifnum #1\expandafter \@firstoftwo
 \else \expandafter \@secondoftwo
 \fi
}%
\providecommand \@ifx [1]{%
 \ifx #1\expandafter \@firstoftwo
 \else \expandafter \@secondoftwo
 \fi
}%
\providecommand \natexlab [1]{#1}%
\providecommand \enquote  [1]{``#1''}%
\providecommand \bibnamefont  [1]{#1}%
\providecommand \bibfnamefont [1]{#1}%
\providecommand \citenamefont [1]{#1}%
\providecommand \href@noop [0]{\@secondoftwo}%
\providecommand \href [0]{\begingroup \@sanitize@url \@href}%
\providecommand \@href[1]{\@@startlink{#1}\@@href}%
\providecommand \@@href[1]{\endgroup#1\@@endlink}%
\providecommand \@sanitize@url [0]{\catcode `\\12\catcode `\$12\catcode
  `\&12\catcode `\#12\catcode `\^12\catcode `\_12\catcode `\%12\relax}%
\providecommand \@@startlink[1]{}%
\providecommand \@@endlink[0]{}%
\providecommand \url  [0]{\begingroup\@sanitize@url \@url }%
\providecommand \@url [1]{\endgroup\@href {#1}{\urlprefix }}%
\providecommand \urlprefix  [0]{URL }%
\providecommand \Eprint [0]{\href }%
\providecommand \doibase [0]{http://dx.doi.org/}%
\providecommand \selectlanguage [0]{\@gobble}%
\providecommand \bibinfo  [0]{\@secondoftwo}%
\providecommand \bibfield  [0]{\@secondoftwo}%
\providecommand \translation [1]{[#1]}%
\providecommand \BibitemOpen [0]{}%
\providecommand \bibitemStop [0]{}%
\providecommand \bibitemNoStop [0]{.\EOS\space}%
\providecommand \EOS [0]{\spacefactor3000\relax}%
\providecommand \BibitemShut  [1]{\csname bibitem#1\endcsname}%
\let\auto@bib@innerbib\@empty
\bibitem [{\citenamefont {Nienhuis}(1982)}]{nie82}%
  \BibitemOpen
  \bibfield  {author} {\bibinfo {author} {\bibfnamefont {B.}~\bibnamefont
  {Nienhuis}},\ }\href {\doibase 10.1103/PhysRevLett.49.1062} {\bibfield
  {journal} {\bibinfo  {journal} {Phys. Rev. Lett.}\ }\textbf {\bibinfo
  {volume} {49}},\ \bibinfo {pages} {1062} (\bibinfo {year}
  {1982})}\BibitemShut {NoStop}%
\bibitem [{\citenamefont {Di~Francesco}\ \emph {et~al.}(1987)\citenamefont
  {Di~Francesco}, \citenamefont {Saleur},\ and\ \citenamefont {Zuber}}]{fsz87}%
  \BibitemOpen
  \bibfield  {author} {\bibinfo {author} {\bibfnamefont {P.}~\bibnamefont
  {Di~Francesco}}, \bibinfo {author} {\bibfnamefont {H.}~\bibnamefont
  {Saleur}}, \ and\ \bibinfo {author} {\bibfnamefont {J.-B.}\ \bibnamefont
  {Zuber}},\ }\href {\doibase 10.1007/BF01009954} {\bibfield  {journal}
  {\bibinfo  {journal} {Journal of statistical physics}\ }\textbf {\bibinfo
  {volume} {49}},\ \bibinfo {pages} {57} (\bibinfo {year} {1987})}\BibitemShut
  {NoStop}%
\bibitem [{\citenamefont {Belavin}\ \emph {et~al.}(1984)\citenamefont
  {Belavin}, \citenamefont {Polyakov},\ and\ \citenamefont
  {Zamolodchikov}}]{belavin1984infinite}%
  \BibitemOpen
  \bibfield  {author} {\bibinfo {author} {\bibfnamefont {A.~A.}\ \bibnamefont
  {Belavin}}, \bibinfo {author} {\bibfnamefont {A.~M.}\ \bibnamefont
  {Polyakov}}, \ and\ \bibinfo {author} {\bibfnamefont {A.~B.}\ \bibnamefont
  {Zamolodchikov}},\ }\href@noop {} {\bibfield  {journal} {\bibinfo  {journal}
  {Nuclear Physics B}\ }\textbf {\bibinfo {volume} {241}},\ \bibinfo {pages}
  {333} (\bibinfo {year} {1984})}\BibitemShut {NoStop}%
\bibitem [{\citenamefont {Francesco}\ \emph {et~al.}(2012)\citenamefont
  {Francesco}, \citenamefont {Mathieu},\ and\ \citenamefont
  {S{\'e}n{\'e}chal}}]{francesco2012conformal}%
  \BibitemOpen
  \bibfield  {author} {\bibinfo {author} {\bibfnamefont {P.}~\bibnamefont
  {Francesco}}, \bibinfo {author} {\bibfnamefont {P.}~\bibnamefont {Mathieu}},
  \ and\ \bibinfo {author} {\bibfnamefont {D.}~\bibnamefont
  {S{\'e}n{\'e}chal}},\ }\href@noop {} {\emph {\bibinfo {title} {Conformal
  field theory}}}\ (\bibinfo  {publisher} {Springer Science \& Business
  Media},\ \bibinfo {year} {2012})\BibitemShut {NoStop}%
\bibitem [{\citenamefont {{Miller}}\ \emph {et~al.}(2021)\citenamefont
  {{Miller}}, \citenamefont {{Sheffield}},\ and\ \citenamefont
  {{Werner}}}]{msw-non-simple}%
  \BibitemOpen
  \bibfield  {author} {\bibinfo {author} {\bibfnamefont {J.}~\bibnamefont
  {{Miller}}}, \bibinfo {author} {\bibfnamefont {S.}~\bibnamefont
  {{Sheffield}}}, \ and\ \bibinfo {author} {\bibfnamefont {W.}~\bibnamefont
  {{Werner}}},\ }\href@noop {} {\bibfield  {journal} {\bibinfo  {journal}
  {Probab. Theory Relat. Fields}\ }\textbf {\bibinfo {volume} {181}},\ \bibinfo
  {pages} {669} (\bibinfo {year} {2021})}\BibitemShut {NoStop}%
\bibitem [{\citenamefont {{M}iller}\ \emph {et~al.}(2022)\citenamefont
  {{M}iller}, \citenamefont {{S}heffield},\ and\ \citenamefont
  {{W}erner}}]{msw-cle-lqg}%
  \BibitemOpen
  \bibfield  {author} {\bibinfo {author} {\bibfnamefont {J.}~\bibnamefont
  {{M}iller}}, \bibinfo {author} {\bibfnamefont {S.}~\bibnamefont
  {{S}heffield}}, \ and\ \bibinfo {author} {\bibfnamefont {W.}~\bibnamefont
  {{W}erner}},\ }\href@noop {} {\bibfield  {journal} {\bibinfo  {journal} {Ann.
  Probab.}\ }\textbf {\bibinfo {volume} {50}},\ \bibinfo {pages} {905}
  (\bibinfo {year} {2022})}\BibitemShut {NoStop}%
\bibitem [{\citenamefont {Cardy}(1992)}]{cardy1992critical}%
  \BibitemOpen
  \bibfield  {author} {\bibinfo {author} {\bibfnamefont {J.~L.}\ \bibnamefont
  {Cardy}},\ }\href@noop {} {\bibfield  {journal} {\bibinfo  {journal} {Journal
  of Physics A: Mathematical and General}\ }\textbf {\bibinfo {volume} {25}},\
  \bibinfo {pages} {L201} (\bibinfo {year} {1992})}\BibitemShut {NoStop}%
\bibitem [{\citenamefont {Nolin}\ \emph {et~al.}(2025)\citenamefont {Nolin},
  \citenamefont {Qian}, \citenamefont {Sun},\ and\ \citenamefont
  {Zhuang}}]{NQSZ-backbone}%
  \BibitemOpen
  \bibfield  {author} {\bibinfo {author} {\bibfnamefont {P.}~\bibnamefont
  {Nolin}}, \bibinfo {author} {\bibfnamefont {W.}~\bibnamefont {Qian}},
  \bibinfo {author} {\bibfnamefont {X.}~\bibnamefont {Sun}}, \ and\ \bibinfo
  {author} {\bibfnamefont {Z.}~\bibnamefont {Zhuang}},\ }\href {\doibase
  10.1103/physrevlett.134.117101} {\bibfield  {journal} {\bibinfo  {journal}
  {Phys. Rev. Lett.}\ }\textbf {\bibinfo {volume} {134}},\ \bibinfo {pages}
  {Paper No. 117101, 6} (\bibinfo {year} {2025})}\BibitemShut {NoStop}%
\bibitem [{\citenamefont {Cardy}(2006)}]{Car06}%
  \BibitemOpen
  \bibfield  {author} {\bibinfo {author} {\bibfnamefont {J.}~\bibnamefont
  {Cardy}},\ }\href {\doibase 10.1007/s10955-006-9186-8} {\bibfield  {journal}
  {\bibinfo  {journal} {J. Stat. Phys.}\ }\textbf {\bibinfo {volume} {125}},\
  \bibinfo {pages} {1} (\bibinfo {year} {2006})}\BibitemShut {NoStop}%
\bibitem [{\citenamefont {Dubail}\ \emph {et~al.}(2009)\citenamefont {Dubail},
  \citenamefont {Jacobsen},\ and\ \citenamefont {Saleur}}]{DJS2009}%
  \BibitemOpen
  \bibfield  {author} {\bibinfo {author} {\bibfnamefont {J.}~\bibnamefont
  {Dubail}}, \bibinfo {author} {\bibfnamefont {J.~L.}\ \bibnamefont
  {Jacobsen}}, \ and\ \bibinfo {author} {\bibfnamefont {H.}~\bibnamefont
  {Saleur}},\ }\href {\doibase 10.1016/j.nuclphysb.2008.12.023} {\bibfield
  {journal} {\bibinfo  {journal} {Nuclear Phys. B}\ }\textbf {\bibinfo {volume}
  {813}},\ \bibinfo {pages} {430} (\bibinfo {year} {2009})}\BibitemShut
  {NoStop}%
\bibitem [{\citenamefont {Grans-Samuelsson}\ \emph {et~al.}(2023)\citenamefont
  {Grans-Samuelsson}, \citenamefont {Jacobsen}, \citenamefont {Nivesvivat},
  \citenamefont {Ribault},\ and\ \citenamefont {Saleur}}]{gjnrs23}%
  \BibitemOpen
  \bibfield  {author} {\bibinfo {author} {\bibfnamefont {L.}~\bibnamefont
  {Grans-Samuelsson}}, \bibinfo {author} {\bibfnamefont {J.~L.}\ \bibnamefont
  {Jacobsen}}, \bibinfo {author} {\bibfnamefont {R.}~\bibnamefont
  {Nivesvivat}}, \bibinfo {author} {\bibfnamefont {S.}~\bibnamefont {Ribault}},
  \ and\ \bibinfo {author} {\bibfnamefont {H.}~\bibnamefont {Saleur}},\ }\href
  {\doibase 10.21468/SciPostPhys.15.4.147} {\bibfield  {journal} {\bibinfo
  {journal} {SciPost Phys.}\ }\textbf {\bibinfo {volume} {15}},\ \bibinfo
  {pages} {147} (\bibinfo {year} {2023})},\ \Eprint
  {http://arxiv.org/abs/2302.08168} {arXiv:2302.08168 [hep-th]} \BibitemShut
  {NoStop}%
\bibitem [{\citenamefont {Grans-Samuelsson}\ \emph {et~al.}(2022)\citenamefont
  {Grans-Samuelsson}, \citenamefont {Nivesvivat}, \citenamefont {Jacobsen},
  \citenamefont {Ribault},\ and\ \citenamefont {Saleur}}]{gnjrs21}%
  \BibitemOpen
  \bibfield  {author} {\bibinfo {author} {\bibfnamefont {L.}~\bibnamefont
  {Grans-Samuelsson}}, \bibinfo {author} {\bibfnamefont {R.}~\bibnamefont
  {Nivesvivat}}, \bibinfo {author} {\bibfnamefont {J.~L.}\ \bibnamefont
  {Jacobsen}}, \bibinfo {author} {\bibfnamefont {S.}~\bibnamefont {Ribault}}, \
  and\ \bibinfo {author} {\bibfnamefont {H.}~\bibnamefont {Saleur}},\ }\href
  {\doibase 10.21468/SciPostPhys.12.5.147} {\bibfield  {journal} {\bibinfo
  {journal} {{SciPost Phys.}}\ }\textbf {\bibinfo {volume} {12}},\ \bibinfo
  {pages} {147} (\bibinfo {year} {2022})},\ \Eprint
  {http://arxiv.org/abs/2111.01106} {arXiv:2111.01106 [hep-th]} \BibitemShut
  {NoStop}%
\bibitem [{\citenamefont {Jacobsen}\ \emph {et~al.}(2025)\citenamefont
  {Jacobsen}, \citenamefont {Nivesvivat}, \citenamefont {Ribault},\ and\
  \citenamefont {Roux}}]{JNRR25}%
  \BibitemOpen
  \bibfield  {author} {\bibinfo {author} {\bibfnamefont {J.~L.}\ \bibnamefont
  {Jacobsen}}, \bibinfo {author} {\bibfnamefont {R.}~\bibnamefont
  {Nivesvivat}}, \bibinfo {author} {\bibfnamefont {S.}~\bibnamefont {Ribault}},
  \ and\ \bibinfo {author} {\bibfnamefont {P.}~\bibnamefont {Roux}},\
  }\href@noop {} {\bibfield  {journal} {\bibinfo  {journal} {arXiv preprint}\ }
  (\bibinfo {year} {2025})},\ \bibinfo {note} {arXiv:2510.04701},\ \Eprint
  {http://arxiv.org/abs/2510.04701} {arXiv:2510.04701 [hep-th]} \BibitemShut
  {NoStop}%
\bibitem [{\citenamefont {Ang}\ \emph {et~al.}(2024{\natexlab{a}})\citenamefont
  {Ang}, \citenamefont {Cai}, \citenamefont {Sun},\ and\ \citenamefont
  {Wu}}]{acsw21}%
  \BibitemOpen
  \bibfield  {author} {\bibinfo {author} {\bibfnamefont {M.}~\bibnamefont
  {Ang}}, \bibinfo {author} {\bibfnamefont {G.}~\bibnamefont {Cai}}, \bibinfo
  {author} {\bibfnamefont {X.}~\bibnamefont {Sun}}, \ and\ \bibinfo {author}
  {\bibfnamefont {B.}~\bibnamefont {Wu}},\ }\href@noop {} {\enquote {\bibinfo
  {title} {{Integrability of Conformal Loop Ensemble: Imaginary DOZZ Formula
  and Beyond}},}\ } (\bibinfo {year} {2024}{\natexlab{a}}),\ \Eprint
  {http://arxiv.org/abs/2107.01788} {arXiv:2107.01788 [math-ph]} \BibitemShut
  {NoStop}%
\bibitem [{\citenamefont {Nivesvivat}\ \emph {et~al.}(2024)\citenamefont
  {Nivesvivat}, \citenamefont {Ribault},\ and\ \citenamefont
  {Jacobsen}}]{nrj23}%
  \BibitemOpen
  \bibfield  {author} {\bibinfo {author} {\bibfnamefont {R.}~\bibnamefont
  {Nivesvivat}}, \bibinfo {author} {\bibfnamefont {S.}~\bibnamefont {Ribault}},
  \ and\ \bibinfo {author} {\bibfnamefont {J.~L.}\ \bibnamefont {Jacobsen}},\
  }\href@noop {} {\bibfield  {journal} {\bibinfo  {journal} {SciPost Physics}\
  }\textbf {\bibinfo {volume} {17}},\ \bibinfo {pages} {029} (\bibinfo {year}
  {2024})}\BibitemShut {NoStop}%
\bibitem [{\citenamefont {Dotsenko}\ and\ \citenamefont {Fateev}(1984)}]{DF84}%
  \BibitemOpen
  \bibfield  {author} {\bibinfo {author} {\bibfnamefont {V.}~\bibnamefont
  {Dotsenko}}\ and\ \bibinfo {author} {\bibfnamefont {V.}~\bibnamefont
  {Fateev}},\ }\href {\doibase https://doi.org/10.1016/0550-3213(84)90269-4}
  {\bibfield  {journal} {\bibinfo  {journal} {Nuclear Physics B}\ }\textbf
  {\bibinfo {volume} {240}},\ \bibinfo {pages} {312} (\bibinfo {year}
  {1984})}\BibitemShut {NoStop}%
\bibitem [{\citenamefont {Delfino}\ and\ \citenamefont
  {Viti}(2010)}]{delfino2010three}%
  \BibitemOpen
  \bibfield  {author} {\bibinfo {author} {\bibfnamefont {G.}~\bibnamefont
  {Delfino}}\ and\ \bibinfo {author} {\bibfnamefont {J.}~\bibnamefont {Viti}},\
  }\href@noop {} {\bibfield  {journal} {\bibinfo  {journal} {Journal of Physics
  A: Mathematical and Theoretical}\ }\textbf {\bibinfo {volume} {44}},\
  \bibinfo {pages} {032001} (\bibinfo {year} {2010})}\BibitemShut {NoStop}%
\bibitem [{\citenamefont {Ikhlef}\ \emph {et~al.}(2016)\citenamefont {Ikhlef},
  \citenamefont {Jacobsen},\ and\ \citenamefont {Saleur}}]{ikhlef2016three}%
  \BibitemOpen
  \bibfield  {author} {\bibinfo {author} {\bibfnamefont {Y.}~\bibnamefont
  {Ikhlef}}, \bibinfo {author} {\bibfnamefont {J.~L.}\ \bibnamefont
  {Jacobsen}}, \ and\ \bibinfo {author} {\bibfnamefont {H.}~\bibnamefont
  {Saleur}},\ }\href@noop {} {\bibfield  {journal} {\bibinfo  {journal}
  {Physical review letters}\ }\textbf {\bibinfo {volume} {116}},\ \bibinfo
  {pages} {130601} (\bibinfo {year} {2016})}\BibitemShut {NoStop}%
\bibitem [{\citenamefont {Schomerus}(2003)}]{Schomerus2003}%
  \BibitemOpen
  \bibfield  {author} {\bibinfo {author} {\bibfnamefont {V.}~\bibnamefont
  {Schomerus}},\ }\href {\doibase 10.1088/1126-6708/2003/11/043} {\bibfield
  {journal} {\bibinfo  {journal} {J. High Energy Phys.}\ ,\ \bibinfo {pages}
  {043, 19}} (\bibinfo {year} {2003})}\BibitemShut {NoStop}%
\bibitem [{\citenamefont {Zamolodchikov}(2005)}]{zamolodchikov2005three}%
  \BibitemOpen
  \bibfield  {author} {\bibinfo {author} {\bibfnamefont {A.~B.}\ \bibnamefont
  {Zamolodchikov}},\ }\href@noop {} {\bibfield  {journal} {\bibinfo  {journal}
  {Theoretical and mathematical physics}\ }\textbf {\bibinfo {volume} {142}},\
  \bibinfo {pages} {183} (\bibinfo {year} {2005})}\BibitemShut {NoStop}%
\bibitem [{\citenamefont {Kostov}\ and\ \citenamefont
  {Petkova}(2007)}]{Kostov-DOZZ}%
  \BibitemOpen
  \bibfield  {author} {\bibinfo {author} {\bibfnamefont {I.~K.}\ \bibnamefont
  {Kostov}}\ and\ \bibinfo {author} {\bibfnamefont {V.~B.}\ \bibnamefont
  {Petkova}},\ }\href {\doibase 10.1016/j.nuclphysb.2007.02.014} {\bibfield
  {journal} {\bibinfo  {journal} {Nucl. Phys. B}\ }\textbf {\bibinfo {volume}
  {770}},\ \bibinfo {pages} {273} (\bibinfo {year} {2007})},\ \Eprint
  {http://arxiv.org/abs/hep-th/0512346} {arXiv:hep-th/0512346} \BibitemShut
  {NoStop}%
\bibitem [{\citenamefont {{Kupiainen}}\ \emph {et~al.}(2020)\citenamefont
  {{Kupiainen}}, \citenamefont {{Rhodes}},\ and\ \citenamefont
  {{Vargas}}}]{krv-dozz}%
  \BibitemOpen
  \bibfield  {author} {\bibinfo {author} {\bibfnamefont {A.}~\bibnamefont
  {{Kupiainen}}}, \bibinfo {author} {\bibfnamefont {R.}~\bibnamefont
  {{Rhodes}}}, \ and\ \bibinfo {author} {\bibfnamefont {V.}~\bibnamefont
  {{Vargas}}},\ }\href@noop {} {\bibfield  {journal} {\bibinfo  {journal} {Ann.
  Math.}\ }\textbf {\bibinfo {volume} {191}},\ \bibinfo {pages} {81} (\bibinfo
  {year} {2020})}\BibitemShut {NoStop}%
\bibitem [{\citenamefont {Ang}\ \emph {et~al.}(2025)\citenamefont {Ang},
  \citenamefont {Cai}, \citenamefont {Jacobsen}, \citenamefont {Nivesvivat},
  \citenamefont {Roux}, \citenamefont {Sun},\ and\ \citenamefont {Wu}}]{sm}%
  \BibitemOpen
  \bibfield  {author} {\bibinfo {author} {\bibfnamefont {M.}~\bibnamefont
  {Ang}}, \bibinfo {author} {\bibfnamefont {G.}~\bibnamefont {Cai}}, \bibinfo
  {author} {\bibfnamefont {J.}~\bibnamefont {Jacobsen}}, \bibinfo {author}
  {\bibfnamefont {R.}~\bibnamefont {Nivesvivat}}, \bibinfo {author}
  {\bibfnamefont {P.}~\bibnamefont {Roux}}, \bibinfo {author} {\bibfnamefont
  {X.}~\bibnamefont {Sun}}, \ and\ \bibinfo {author} {\bibfnamefont
  {B.}~\bibnamefont {Wu}},\ }\href@noop {} {\enquote {\bibinfo {title}
  {Supplemental material},}\ } (\bibinfo {year} {2025}),\ \bibinfo {note} {see
  Supplemental Material for definitions and technical details}\BibitemShut
  {NoStop}%
\bibitem [{\citenamefont {He}\ \emph {et~al.}(2020)\citenamefont {He},
  \citenamefont {Jacobsen},\ and\ \citenamefont {Saleur}}]{hjs20}%
  \BibitemOpen
  \bibfield  {author} {\bibinfo {author} {\bibfnamefont {Y.}~\bibnamefont
  {He}}, \bibinfo {author} {\bibfnamefont {J.~L.}\ \bibnamefont {Jacobsen}}, \
  and\ \bibinfo {author} {\bibfnamefont {H.}~\bibnamefont {Saleur}},\ }\href
  {\doibase 10.1007/JHEP12(2020)019} {\  (\bibinfo {year} {2020}),\
  10.1007/JHEP12(2020)019},\ \Eprint {http://arxiv.org/abs/2005.07258}
  {arXiv:2005.07258 [hep-th]} \BibitemShut {NoStop}%
\bibitem [{\citenamefont {Bl\"ote}\ and\ \citenamefont {Nienhuis}()}]{BN1989}%
  \BibitemOpen
  \bibfield  {author} {\bibinfo {author} {\bibfnamefont {H.~W.~J.}\
  \bibnamefont {Bl\"ote}}\ and\ \bibinfo {author} {\bibfnamefont
  {B.}~\bibnamefont {Nienhuis}},\ }\href {\doibase 10.1088/0305-4470/22/9/028}
  {\bibfield  {journal} {\bibinfo  {journal} {J. Phys. A: Math. Gen.}\ }\textbf
  {\bibinfo {volume} {22}},\ \bibinfo {pages} {1415}}\BibitemShut {NoStop}%
\bibitem [{\citenamefont {Nivesvivat}\ and\ \citenamefont
  {Ribault}(2021)}]{nr20}%
  \BibitemOpen
  \bibfield  {author} {\bibinfo {author} {\bibfnamefont {R.}~\bibnamefont
  {Nivesvivat}}\ and\ \bibinfo {author} {\bibfnamefont {S.}~\bibnamefont
  {Ribault}},\ }\href {\doibase 10.21468/SciPostPhys.10.1.021} {\bibfield
  {journal} {\bibinfo  {journal} {SciPost Phys.}\ }\textbf {\bibinfo {volume}
  {10}},\ \bibinfo {pages} {021} (\bibinfo {year} {2021})},\ \Eprint
  {http://arxiv.org/abs/2007.04190} {arXiv:2007.04190 [hep-th]} \BibitemShut
  {NoStop}%
\bibitem [{\citenamefont {Migliaccio}\ and\ \citenamefont
  {Ribault}(2018)}]{mr17}%
  \BibitemOpen
  \bibfield  {author} {\bibinfo {author} {\bibfnamefont {S.}~\bibnamefont
  {Migliaccio}}\ and\ \bibinfo {author} {\bibfnamefont {S.}~\bibnamefont
  {Ribault}},\ }\href {\doibase 10.1007/JHEP05(2018)169} {\bibfield  {journal}
  {\bibinfo  {journal} {JHEP}\ }\textbf {\bibinfo {volume} {05}},\ \bibinfo
  {pages} {169} (\bibinfo {year} {2018})},\ \Eprint
  {http://arxiv.org/abs/1711.08916} {arXiv:1711.08916 [hep-th]} \BibitemShut
  {NoStop}%
\bibitem [{\citenamefont {Camia}\ and\ \citenamefont
  {Newman}(2006)}]{CN06-full}%
  \BibitemOpen
  \bibfield  {author} {\bibinfo {author} {\bibfnamefont {F.}~\bibnamefont
  {Camia}}\ and\ \bibinfo {author} {\bibfnamefont {C.~M.}\ \bibnamefont
  {Newman}},\ }\href {\doibase 10.1007/s00220-006-0086-1} {\bibfield  {journal}
  {\bibinfo  {journal} {Comm. Math. Phys.}\ }\textbf {\bibinfo {volume}
  {268}},\ \bibinfo {pages} {1} (\bibinfo {year} {2006})}\BibitemShut {NoStop}%
\bibitem [{\citenamefont {Sheffield}(2009)}]{SheffieldCLE}%
  \BibitemOpen
  \bibfield  {author} {\bibinfo {author} {\bibfnamefont {S.}~\bibnamefont
  {Sheffield}},\ }\href {\doibase 10.1215/00127094-2009-007} {\bibfield
  {journal} {\bibinfo  {journal} {Duke Math. J.}\ }\textbf {\bibinfo {volume}
  {147}},\ \bibinfo {pages} {79} (\bibinfo {year} {2009})}\BibitemShut
  {NoStop}%
\bibitem [{\citenamefont {Schramm}(2000)}]{Sc00}%
  \BibitemOpen
  \bibfield  {author} {\bibinfo {author} {\bibfnamefont {O.}~\bibnamefont
  {Schramm}},\ }\href {\doibase 10.1007/BF02803524} {\bibfield  {journal}
  {\bibinfo  {journal} {Israel J. Math.}\ }\textbf {\bibinfo {volume} {118}},\
  \bibinfo {pages} {221} (\bibinfo {year} {2000})}\BibitemShut {NoStop}%
\bibitem [{\citenamefont {Smirnov}(2001)}]{Sm01}%
  \BibitemOpen
  \bibfield  {author} {\bibinfo {author} {\bibfnamefont {S.}~\bibnamefont
  {Smirnov}},\ }\href {\doibase 10.1016/S0764-4442(01)01991-7} {\bibfield
  {journal} {\bibinfo  {journal} {C. R. Acad. Sci. Paris S\'{e}r. I Math.}\
  }\textbf {\bibinfo {volume} {333}},\ \bibinfo {pages} {239} (\bibinfo {year}
  {2001})}\BibitemShut {NoStop}%
\bibitem [{\citenamefont {Garban}\ \emph {et~al.}(2013)\citenamefont {Garban},
  \citenamefont {Pete},\ and\ \citenamefont {Schramm}}]{GPS2013}%
  \BibitemOpen
  \bibfield  {author} {\bibinfo {author} {\bibfnamefont {C.}~\bibnamefont
  {Garban}}, \bibinfo {author} {\bibfnamefont {G.}~\bibnamefont {Pete}}, \ and\
  \bibinfo {author} {\bibfnamefont {O.}~\bibnamefont {Schramm}},\ }\href
  {\doibase 10.1090/S0894-0347-2013-00772-9} {\bibfield  {journal} {\bibinfo
  {journal} {J. Amer. Math. Soc.}\ }\textbf {\bibinfo {volume} {26}},\ \bibinfo
  {pages} {939} (\bibinfo {year} {2013})}\BibitemShut {NoStop}%
\bibitem [{\citenamefont {Duplantier}\ and\ \citenamefont
  {Sheffield}(2011)}]{DS-KPZ-invent}%
  \BibitemOpen
  \bibfield  {author} {\bibinfo {author} {\bibfnamefont {B.}~\bibnamefont
  {Duplantier}}\ and\ \bibinfo {author} {\bibfnamefont {S.}~\bibnamefont
  {Sheffield}},\ }\href {\doibase 10.1007/s00222-010-0308-1} {\bibfield
  {journal} {\bibinfo  {journal} {Invent. Math.}\ }\textbf {\bibinfo {volume}
  {185}},\ \bibinfo {pages} {333} (\bibinfo {year} {2011})}\BibitemShut
  {NoStop}%
\bibitem [{\citenamefont {Sheffield}(2016)}]{She16a}%
  \BibitemOpen
  \bibfield  {author} {\bibinfo {author} {\bibfnamefont {S.}~\bibnamefont
  {Sheffield}},\ }\href {\doibase 10.1214/15-AOP1055} {\bibfield  {journal}
  {\bibinfo  {journal} {Ann. Probab.}\ }\textbf {\bibinfo {volume} {44}},\
  \bibinfo {pages} {3474} (\bibinfo {year} {2016})}\BibitemShut {NoStop}%
\bibitem [{\citenamefont {Duplantier}\ \emph {et~al.}(2021)\citenamefont
  {Duplantier}, \citenamefont {Miller},\ and\ \citenamefont
  {Sheffield}}]{DMS14}%
  \BibitemOpen
  \bibfield  {author} {\bibinfo {author} {\bibfnamefont {B.}~\bibnamefont
  {Duplantier}}, \bibinfo {author} {\bibfnamefont {J.}~\bibnamefont {Miller}},
  \ and\ \bibinfo {author} {\bibfnamefont {S.}~\bibnamefont {Sheffield}},\
  }\href {\doibase 10.24033/ast} {\bibfield  {journal} {\bibinfo  {journal}
  {Ast\'erisque}\ ,\ \bibinfo {pages} {viii+257}} (\bibinfo {year}
  {2021})}\BibitemShut {NoStop}%
\bibitem [{\citenamefont {Dorn}\ and\ \citenamefont {Otto}(1994)}]{DO94}%
  \BibitemOpen
  \bibfield  {author} {\bibinfo {author} {\bibfnamefont {H.}~\bibnamefont
  {Dorn}}\ and\ \bibinfo {author} {\bibfnamefont {H.-J.}\ \bibnamefont
  {Otto}},\ }\href {\doibase 10.1016/0550-3213(94)00352-1} {\bibfield
  {journal} {\bibinfo  {journal} {Nuclear Phys. B}\ }\textbf {\bibinfo {volume}
  {429}},\ \bibinfo {pages} {375} (\bibinfo {year} {1994})}\BibitemShut
  {NoStop}%
\bibitem [{\citenamefont {Zamolodchikov}\ and\ \citenamefont
  {Zamolodchikov}(1996)}]{ZZ96}%
  \BibitemOpen
  \bibfield  {author} {\bibinfo {author} {\bibfnamefont {A.~B.}\ \bibnamefont
  {Zamolodchikov}}\ and\ \bibinfo {author} {\bibfnamefont {A.~B.}\ \bibnamefont
  {Zamolodchikov}},\ }\href {\doibase 10.1016/0550-3213(96)00351-3} {\bibfield
  {journal} {\bibinfo  {journal} {Nucl. Phys. B}\ }\textbf {\bibinfo {volume}
  {477}},\ \bibinfo {pages} {577} (\bibinfo {year} {1996})},\ \Eprint
  {http://arxiv.org/abs/hep-th/9506136} {arXiv:hep-th/9506136} \BibitemShut
  {NoStop}%
\bibitem [{\citenamefont {Ang}\ \emph {et~al.}(2024{\natexlab{b}})\citenamefont
  {Ang}, \citenamefont {Holden},\ and\ \citenamefont
  {Sun}}]{AHS-SLE-integrability}%
  \BibitemOpen
  \bibfield  {author} {\bibinfo {author} {\bibfnamefont {M.}~\bibnamefont
  {Ang}}, \bibinfo {author} {\bibfnamefont {N.}~\bibnamefont {Holden}}, \ and\
  \bibinfo {author} {\bibfnamefont {X.}~\bibnamefont {Sun}},\ }\href {\doibase
  10.1002/cpa.22180} {\bibfield  {journal} {\bibinfo  {journal} {Comm. Pure
  Appl. Math.}\ }\textbf {\bibinfo {volume} {77}},\ \bibinfo {pages} {2651}
  (\bibinfo {year} {2024}{\natexlab{b}})}\BibitemShut {NoStop}%
\bibitem [{\citenamefont {Nolin}\ \emph {et~al.}(2023)\citenamefont {Nolin},
  \citenamefont {Qian}, \citenamefont {Sun},\ and\ \citenamefont
  {Zhuang}}]{nolin2024}%
  \BibitemOpen
  \bibfield  {author} {\bibinfo {author} {\bibfnamefont {P.}~\bibnamefont
  {Nolin}}, \bibinfo {author} {\bibfnamefont {W.}~\bibnamefont {Qian}},
  \bibinfo {author} {\bibfnamefont {X.}~\bibnamefont {Sun}}, \ and\ \bibinfo
  {author} {\bibfnamefont {Z.}~\bibnamefont {Zhuang}},\ }\href@noop {}
  {\bibfield  {journal} {\bibinfo  {journal} {arXiv preprint arXiv:2309.05050}\
  } (\bibinfo {year} {2023})}\BibitemShut {NoStop}%
\bibitem [{\citenamefont {Ang}\ \emph {et~al.}(2024{\natexlab{c}})\citenamefont
  {Ang}, \citenamefont {Sun}, \citenamefont {Yu},\ and\ \citenamefont
  {Zhuang}}]{ang2024boundary}%
  \BibitemOpen
  \bibfield  {author} {\bibinfo {author} {\bibfnamefont {M.}~\bibnamefont
  {Ang}}, \bibinfo {author} {\bibfnamefont {X.}~\bibnamefont {Sun}}, \bibinfo
  {author} {\bibfnamefont {P.}~\bibnamefont {Yu}}, \ and\ \bibinfo {author}
  {\bibfnamefont {Z.}~\bibnamefont {Zhuang}},\ }\href@noop {} {\bibfield
  {journal} {\bibinfo  {journal} {arXiv preprint arXiv:2401.15904}\ } (\bibinfo
  {year} {2024}{\natexlab{c}})}\BibitemShut {NoStop}%
\bibitem [{\citenamefont {Ang}\ \emph {et~al.}(2022)\citenamefont {Ang},
  \citenamefont {Remy},\ and\ \citenamefont {Sun}}]{ars-annuli}%
  \BibitemOpen
  \bibfield  {author} {\bibinfo {author} {\bibfnamefont {M.}~\bibnamefont
  {Ang}}, \bibinfo {author} {\bibfnamefont {G.}~\bibnamefont {Remy}}, \ and\
  \bibinfo {author} {\bibfnamefont {X.}~\bibnamefont {Sun}},\ }\href@noop {}
  {\bibfield  {journal} {\bibinfo  {journal} {arXiv preprint arXiv:2203.12398}\
  } (\bibinfo {year} {2022})}\BibitemShut {NoStop}%
\bibitem [{\citenamefont {Polyakov}(1981)}]{polyakov1981quantum}%
  \BibitemOpen
  \bibfield  {author} {\bibinfo {author} {\bibfnamefont {A.~M.}\ \bibnamefont
  {Polyakov}},\ }\href {\doibase 10.1016/0370-2693(81)90743-7} {\bibfield
  {journal} {\bibinfo  {journal} {Phys. Lett. B}\ }\textbf {\bibinfo {volume}
  {103}},\ \bibinfo {pages} {207} (\bibinfo {year} {1981})}\BibitemShut
  {NoStop}%
\bibitem [{\citenamefont {David}(1988)}]{david-conformal-gauge}%
  \BibitemOpen
  \bibfield  {author} {\bibinfo {author} {\bibfnamefont {F.}~\bibnamefont
  {David}},\ }\href@noop {} {\bibfield  {journal} {\bibinfo  {journal} {{M}od.
  {P}hys. {L}ett. {A}}\ }\textbf {\bibinfo {volume} {3}},\ \bibinfo {pages}
  {1651} (\bibinfo {year} {1988})}\BibitemShut {NoStop}%
\bibitem [{\citenamefont {Distler}\ and\ \citenamefont {Kawai}(1989)}]{dk-qg}%
  \BibitemOpen
  \bibfield  {author} {\bibinfo {author} {\bibfnamefont {J.}~\bibnamefont
  {Distler}}\ and\ \bibinfo {author} {\bibfnamefont {H.}~\bibnamefont
  {Kawai}},\ }\href@noop {} {\bibfield  {journal} {\bibinfo  {journal}
  {{N}ucl.{P}hys. {B}}\ }\textbf {\bibinfo {volume} {321}},\ \bibinfo {pages}
  {509–527} (\bibinfo {year} {1989})}\BibitemShut {NoStop}%
\bibitem [{\citenamefont {Knizhnik}\ \emph {et~al.}(1988)\citenamefont
  {Knizhnik}, \citenamefont {Polyakov},\ and\ \citenamefont
  {Zamolodchikov}}]{KPZrelation}%
  \BibitemOpen
  \bibfield  {author} {\bibinfo {author} {\bibfnamefont {V.~G.}\ \bibnamefont
  {Knizhnik}}, \bibinfo {author} {\bibfnamefont {A.~M.}\ \bibnamefont
  {Polyakov}}, \ and\ \bibinfo {author} {\bibfnamefont {A.~B.}\ \bibnamefont
  {Zamolodchikov}},\ }\href {\doibase 10.1142/S0217732388000982} {\bibfield
  {journal} {\bibinfo  {journal} {Mod. Phys. Lett. A}\ }\textbf {\bibinfo
  {volume} {3}},\ \bibinfo {pages} {819} (\bibinfo {year} {1988})}\BibitemShut
  {NoStop}%
\bibitem [{\citenamefont {Beffara}(2008)}]{Beffara_2008}%
  \BibitemOpen
  \bibfield  {author} {\bibinfo {author} {\bibfnamefont {V.}~\bibnamefont
  {Beffara}},\ }\href {\doibase 10.1214/07-aop364} {\bibfield  {journal}
  {\bibinfo  {journal} {The Annals of Probability}\ }\textbf {\bibinfo {volume}
  {36}} (\bibinfo {year} {2008}),\ 10.1214/07-aop364}\BibitemShut {NoStop}%
\bibitem [{\citenamefont {Cai}\ \emph {et~al.}(2025)\citenamefont {Cai},
  \citenamefont {Liu}, \citenamefont {Wu},\ and\ \citenamefont
  {Zhuang}}]{CLWZ25}%
  \BibitemOpen
  \bibfield  {author} {\bibinfo {author} {\bibfnamefont {G.}~\bibnamefont
  {Cai}}, \bibinfo {author} {\bibfnamefont {H.}~\bibnamefont {Liu}}, \bibinfo
  {author} {\bibfnamefont {B.}~\bibnamefont {Wu}}, \ and\ \bibinfo {author}
  {\bibfnamefont {Z.}~\bibnamefont {Zhuang}},\ }\href
  {https://arxiv.org/abs/2510.05850} {\enquote {\bibinfo {title} {Three-point
  connectivity constant for $q$-state potts spin clusters},}\ } (\bibinfo
  {year} {2025}),\ \Eprint {http://arxiv.org/abs/2510.05850} {arXiv:2510.05850
  [math.PR]} \BibitemShut {NoStop}%
\bibitem [{\citenamefont {Delfino}\ \emph {et~al.}(2013)\citenamefont
  {Delfino}, \citenamefont {Picco}, \citenamefont {Santachiara},\ and\
  \citenamefont {Viti}}]{DPSV13}%
  \BibitemOpen
  \bibfield  {author} {\bibinfo {author} {\bibfnamefont {G.}~\bibnamefont
  {Delfino}}, \bibinfo {author} {\bibfnamefont {M.}~\bibnamefont {Picco}},
  \bibinfo {author} {\bibfnamefont {R.}~\bibnamefont {Santachiara}}, \ and\
  \bibinfo {author} {\bibfnamefont {J.}~\bibnamefont {Viti}},\ }\href {\doibase
  10.1088/1742-5468/2013/11/p11011} {\bibfield  {journal} {\bibinfo  {journal}
  {Journal of Statistical Mechanics: Theory and Experiment}\ }\textbf {\bibinfo
  {volume} {2013}},\ \bibinfo {pages} {P11011} (\bibinfo {year}
  {2013})}\BibitemShut {NoStop}%
\end{thebibliography}%

\end{document}